\documentclass[conference]{IEEEtran}
\usepackage[utf8]{inputenc}
\usepackage{graphicx}
\usepackage{float}
\usepackage{mathtools}
\graphicspath{ {images/} }
\usepackage[justification=centering, font=normalsize]{caption}
\usepackage[table,xcdraw]{xcolor}
\usepackage{lettrine}
\usepackage{nopageno}
\usepackage{algorithm,algorithmic}
\usepackage{url}
\usepackage{cite}
\usepackage{flushend}

\title{Trajectory Optimization of Flying Energy Sources using Q-Learning to Recharge Hotspot UAVs}

 \author{\IEEEauthorblockN{Sayed Amir Hoseini$^1$,  Jahan Hassan$^2$, Ayub Bokani$^3$,  Salil S. Kanhere$^4$}
\IEEEauthorblockA{$^{1,2,3}\, $School of Engineering and Technology,
The Central Queensland University,
Sydney, Australia \\
\{s.hoseini, j.hassan, a.bokani\}@cqu.edu.au}
$^{4}\, $School of Computer Science and Engineering, 
The University of New South Wales,
Sydney, Australia \\
\ salil.kanhere@unsw.edu.au
}

\begin{document}
\maketitle
\begin{abstract}
Despite the increasing popularity of commercial usage of UAVs or drone-delivered services, their dependence on the limited-capacity on-board batteries hinders their flight-time and mission continuity. As such, developing in-situ power transfer solutions for topping-up UAV batteries have the potential to extend their mission duration. In this paper, we study a scenario where UAVs are deployed as base stations (UAV-BS) providing wireless Hotspot services to the ground nodes, while harvesting wireless energy from flying energy sources. These energy sources are specialized UAVs (Charger or transmitter UAVs, tUAVs), equipped with wireless power transmitting devices such as RF antennae. tUAVs have the flexibility to adjust their flight path to maximize energy transfer. With the increasing number of UAV-BSs and environmental complexity, it is necessary to develop an intelligent trajectory selection procedure for tUAVs so as to optimize the energy transfer gain. In this paper, we model the trajectory optimization of tUAVs as a Markov Decision Process (MDP) problem and solve it using Q-Learning algorithm. Simulation results confirm that the Q-Learning based optimized trajectory of the tUAVs outperforms two benchmark strategies, namely random path planning and static hovering of the tUAVs.
\end{abstract}

\section{Introduction}
The rapid advancement and falling costs of Unmanned Aerial Vehicle (UAVs), a.k.a. drones, are fueling their popularity in a wide range of commercial applications such as goods delivery, medical logistics, entertainment, and aerial imagery\cite{drone_apps}.  The global market of emerging drone-aided commercial services is estimated at a staggering value of $\$127 bn$~\cite{pwc}. Interestingly, the next generation, i.e., 5G and beyond, mobile communication systems are also expected to use UAVs as either aerial base stations/relays/hotspots (UAV-BSs)~\cite{azade_droneBS, Azade_survey}, or as end users using wireless communication services from terrestrial or aerial base stations \cite{Rui_UAV_5g}. For such mobile communication applications, uninterrupted UAV flights is critical to avoid any discontinuity in wireless communications, which may cause mission failure for some applications. Unfortunately, UAV flight-time is limited by the capacity of the on-board battery~\footnote{For example, the typical flight-time of {\it DJI Spreading Wings S900} drone is only $18$ minutes when fully charged \cite{Azade_survey}}, raising a challenge  for uninterrupted UAV missions, especially for smaller commercial drones.   

On-going research to address UAV battery limitation largely concentrated on designing algorithms and motion control functions~\cite{tran2019trajectory,  MICC_ourwork, li2015energy, abdulla2014optimal, zhan2017energy, abdulla2014toward}, that allow UAVs to operate in more energy-efficient way thereby extending the battery life. However, these efforts do not fundamentally solve the problem, because the UAVs still need to leave their missions and return to ground charging stations when the battery ultimately depletes. Utilizing the concept of far-field wireless power transfer (WPT), some researchers have recently contemplated the idea of charging the UAVs through the ground base stations, which eliminates the need for the UAVs to return to a charging station~\cite{RF_UAV, ANSARI2020}. However, the UAVs are required to remain in close proximity of the base stations during the WPT process. Efficient in-situ wireless charging of UAVs therefore remains a challenging open problem. 

Inspired by the mid-air fueling of military jets using aerial tankers~\cite{greener}, in this paper we propose the concept of mid-air UAV wireless charging using specialized drones that are equipped with WPT equipment, henceforth referred to as tUAVs (transmitter UAVs). Under uncertain battery consumption in UAV-BSs due to dynamic mobile communication traffic, we then seek to optimize the trajectory of the flying tUAVs to maximize the long-term utility of the mobile communication service supported by the UAV-BSs. To the best of our knowledge, this is the first attempt to consider such wireless charging concept for UAVs using flying energy sources.     

The main contributions of this paper can be summarized as follows: (i) we propose use of flying transmitter UAVs (tUAVs) to facilitate aerial wireless charging of UAV-BSs; (ii) we show that the trajectory optimization of the tUAVs can be modeled as an MDP given that the movement decision at any given step affects the long-term discounted utility of the underlying mobile communication system supported by the UAV-BSs; (iii) we solve the MDP problem using Q-learning and evaluate the performance of the proposed UAV charging architecture using simulations. Our results confirm that the optimized trajectory outperforms two benchmark strategies, namely random movement and static hovering of the tUAVs. 

 The rest of the paper is organized as follows. The related works are discussed in Section \ref{sec:relwork}. The problem statement and the proposed Q-learning-based trajectory optimization framework is presented in Section \ref{sec:mdp}. Performance evaluation of our proposed framework is discussed in Section \ref{sec:perform}. Finally, we conclude the paper and discuss future works in Section \ref{sec:conclusions}.

\section{Related Work}
\label{sec:relwork}
Energy efficient UAV flight path planning, such as trajectory design methods were proposed in \cite{tran2019trajectory, MICC_ourwork}. While researchers optimize either the mechanical or the electronic energy consumption individually, work reported in \cite{MICC_ourwork} optimizes UAV trajectory  to minimize the energy consumption for the mechanical as well as electronic functions of UAVs. Other researchers proposed a method to reduce the overall energy consumption of UAV communications by extending their network lifetime while guaranteeing their communication's success rate \cite{li2015energy}. Optimal data collection techniques \cite{abdulla2014optimal}, \cite{zhan2017energy} and a UAV-aided networking mechanism \cite{abdulla2014toward} are proposed which positively affect the UAV's energy consumption by optimizing their networking and communication methods. 

Mechanisms for wireless recharging of UAVs have been proposed using terrestrial base stations using RF energy transmission in \cite{RF_UAV} and optical energy transmission in \cite{ANSARI2020}, however, the UAVs are required to stay close to some base stations in order to receive the power. This restricts the locations where the UAVs can be deployed, therefore, efficient in-situ wireless charging of UAVs remains a challenging open problem.

To address this gap, in our previous work \cite{jahan_RF}, we introduced the use of aerial, stationary (i.e., hovering at fixed locations) RF chargers (tUAVs) placed at optimal locations with respect to the serving UAV-BSs (receiver UAVs, rUAVs) for maximizing the total delivered energy. The RF chargers are specialized UAVs carrying RF transmitters and hovering at the same height as the receiver UAV-BSs. We assumed that the tUAVs have sufficient power supply, e.g., by having a larger battery. The transmitter and receiver UAVs' altitude being the same, the placements of the transmitter UAVs were restricted to be outside the collision zone of two UAVs, i.e., the wingspan of separation. We studied their placement optimization to maximize the total harvested energy by the UAV-BSs from the received RF signals. 

Since the received power is dependant on the distance the signals travel as per the well-known Frii's formula \cite{friis-2}, having less or no restriction on the minimum distance between the tUAVs and the UAV-BSs compared to the wingspan distance used in \cite{jahan_RF}, should enhance the received power levels. Moreover, the harvested energy should increase by having the tUAVs moving close to the UAV-BSs reducing the distance further, as opposed to a fixed optimal location. Therefore, building on the above proposal \cite{jahan_RF}, in this paper we introduce the use of \textit{mobile} RF chargers (tUAVs) that  \textit{fly and hover above} the rUAVs (as opposed to having a wingspan of separation distance at the same height of the rUAVs) and transmit wireless energy to extend their flying time. This also achieves direct LoS. The aim of this paper is to study the trajectories of such flying wireless chargers that would maximize the received power by the receiver UAVs. This architecture allows the chargers to carry any type of energy transmitters that can be used for wireless far field energy harvesting, e.g., RF omnidirectional antennae \cite{arrawatia}, massive MIMO with beamforming \cite{wang2018}, Free Space Optics (FSO) \cite{ANSARI2020}, etc. In this paper, we assume  highly directional RF antennae being carried by the tUAV as the source of the energy to increase the efficiency of WPT compared to omnidirectional antennae. We note that despite the specific type of wireless transmitter that is being used, the trajectory optimization of the flying chargers will hold applicable for any type of wireless transmitters due to the impact of distance and Line-of-Sight (LoS) requirement between the transmitter and receiver on the level of received power.


\section{System Description}
\label{sec:mdp}
In this section, we present our UAV recharging architecture, and the Q-Learning formulation to solve the trajectory optimization problem of the tUAVs.


\begin{figure}
\centering 
 {\includegraphics[width=0.48\textwidth]{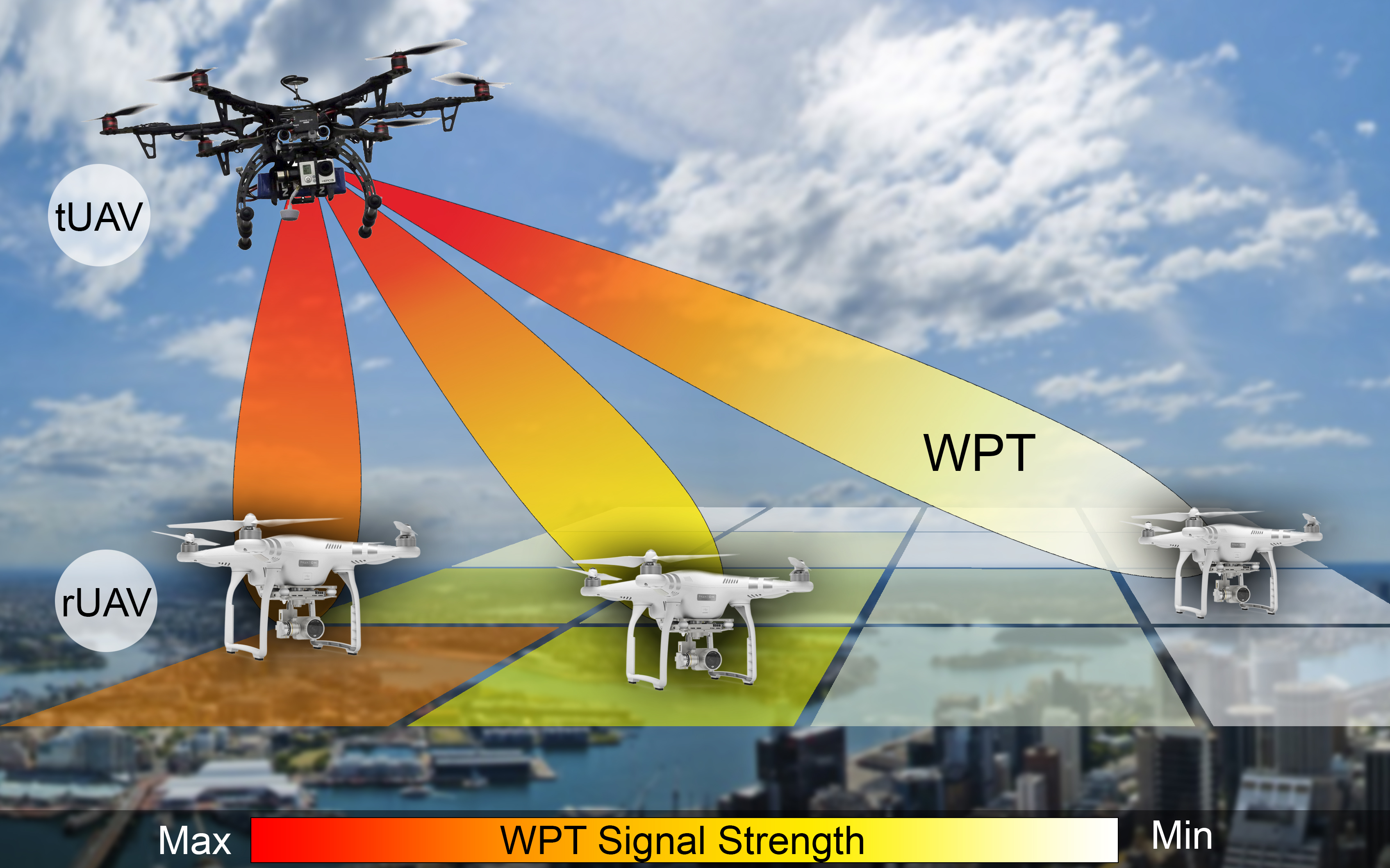}}
\caption{In-situ recharging of UAVs, sample Wireless Power Transfer (WPT) beams in a flying trajectory.}
\label{fig:IDEA} 
\end{figure}

\subsection{UAV Recharging Architecture}

Our proposed UAV recharging architecture consists of specialized, flying UAVs equipped with high gain RF antennae (tUAV) that transmit wireless power to in-situ recharge (or top-up) the batteries of UAV-BSs (rUAVs) that are deployed in an area to provide Hotspots for wireless communication to users/nodes. The architecture is illustrated in Figure~\ref{fig:IDEA}. Our approach offers freedom to address the distance between the transmitters and receivers (tUAVs and rUAVs) that influences wireless energy transfer efficiency. The flying tUAVs are free to fly about and locate themselves such that they can minimize the distance and improve the line-of-sight RF links, as well as address the need of multiple rUAVs for transferring power. This increases the energy transfer utility and efficiency significantly. Unlike the mentioned related work on wireless recharging of UAVs, this architecture allows rUAVs to remain at their deployed locations or trajectories throughout their missions and provide their services while recharging. At the same time, the tUAVs may fly to new locations depending on the movement decision derived from the used algorithms. The decision is made based on rUAVs information which is sent to tUAV periodically. In this paper, we study the use of Q-Learning to make such decisions. Moreover, to formalize the architecture as MDP, we assume that the geographical environment is discretized as a 2D grid where each cell can be covered by one rUAV. The tUAV and rUAVs are located at the center of the cells.

As we have used RF transmitters as the power source, we note that the received power of far-field RF transmission attenuates as per the reciprocal of the squared distance between the transmitter and the receiver. Therefore, assuming full energy conversion efficiency, the harvested RF power ($P_r$) at the rUAVs can be calculated using Frii’s free space propagation model \cite{friis-2} as:

\begin{equation}
    P_r=\frac{P_tG_tG_r\lambda^2}{(4\pi d)^2}
\end{equation}

where $P_t$ is the transmit power, $G_t$ and $G_r$ are the antenna gains of the transmitter and the receiver, $\lambda$ is the power transfer wavelength, and $d$ is the distance between the transmitter and the receiver. 

In particular, we consider that the tUAV is equipped with a few high gain antennae to transfer energy to rUAVs. Since the locations of the UAVs are not being changed rapidly, we assume that each of the tUAV's antennae can be aligned mechanically toward one of the rUAVs. The antenna alignment and high gain can be achieved by MIMO beamforming as an alternative which is not in this paper scope, and left for our future work. Also, a small amount of data is communicated between tUAV and rUAVs periodically using the same wireless channel. This data includes location, battery status and consumed-received energy of each rUAV. Since the RF energy harvesting is limited and may not provide infinite flying time for rUAVs, we assume that the low battery rUAV leaves the network to be charged with other wired or near field WPT methods. However, our solution aims to keep them in operation for as long as possible. 

In this paper, our proposed charging technology is limited to RF power transfer, but the power transfer can be via laser beam and also rUAVs can exploit other in-situ battery charging techniques such as solar energy. We will consider hybrid charging in our future works. We expect our Q-learning approach to be compatible with all of these in-situ charging methods. 


\subsection{Q-Learning Formulation}

Q-learning \cite{Watkins92q-learning} is a model-free reinforcement learning algorithm in which an agent transitions from one state to another, by taking random actions. A set of states $S$ and set of actions $A$ define the learning space. By performing an action $a\in A$ and moving to another state, a \textit{revenue} function calculates a numeric value for taking such state-action pair and records it in a Q-table which is initialized with zero values. By repeatedly taking random actions, at one point the agent reaches a particular goal state. The Q-table values get updated at each step and after many iterations eventually converge. Q-Learning's goal is to maximize the total revenues for all state-action pairs from beginning up to reaching the goal state, so called the \textit{optimal policy}. The optimal policy indicates which action is the best to take in different states, which results in a maximized overall gain.

Q-Learning has been widely used in UAV related research recently. This includes a range of application from military threat avoidance \cite{Yan2019} and obstacle avoidance \cite{yijing2017q} to trajectory optimization for improving services in wireless communications \cite{challita2018deep, bayerlein2018learning}. 

\begin{figure*}
    \centering
    \includegraphics[width=0.75\textwidth]{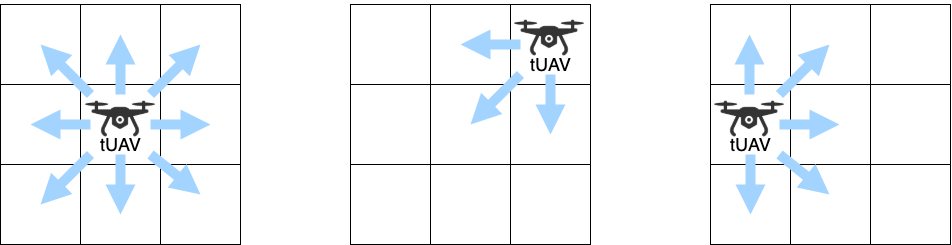}
    \caption{Possible movement directions of the transmitter UAVs (tUAV) with respect to its three example current locations. The tUAV periodically changes its position to improve the energy transfer efficiency. In some periods, the tUAV may find that its current location is the best, thus the position may not be changed.}
    \label{fig:tUAVMove}
\end{figure*}

In this paper, we employ Q-learning to find the best location and movement for the tUAV at a given observation of entire network of rUAVs. Q-Learning components in our solution are defined as following:

\begin{itemize}
    \item \textbf{Agent} (tUAV) observes the current state and takes actions to move to other states.
    
    \item \textbf{Action} ($a$) is defined as flying to a neighboring cell in our assumed grid space, or hovering over the current cell. Hence, as shown in Figure~\ref{fig:tUAVMove}, we consider 9 possible actions. Some actions are not available on the edges of our considered area. 
    
    \item \textbf{State} ($S$) is defined based on the observed information of rUAVs and the current location of  tUAV. Thus, we define the state as $S=\{L_c, L_h, B_h\}$ where $L_c$ is the location of  tUAV, $L_h = [L_{h_1},L_{h_2}, ..., L_{h_n}]$ is a vector that denotes the location of rUAV1 to rUAVn and $B_h = [B_{h_1},B_{h_2}, ..., B_{h_n}]$ is a vector that denotes the battery level of rUAV1 to rUAVn.
    
    \item \textbf{Revenue} ($R$) is a function of last state and action which returns a reward for the energy that all rUAVs receive from tUAV and/or applies a penalty if an rUAV has to move to terrestrial charging station because of low battery. $R$ is formulated as:
    \begin{equation}
        R(S,a) = \mu E_{r,tot} + \nu N_o
    \end{equation}
        where $E_{r,tot}$ is the total harvested energy by rUAVs and $N_o$ is the number of out of charge rUAVs. $\mu$ and $\nu$ are adjusting factors. 
    
\end{itemize}

We consider a time step, $T$ in our model which is roughly long enough for the tUAV to fly from one cell to another. This time step is used to take actions over time, and update the Q-Table values after completing every transition and observing the $current$ state-action pair ($s,a$) as:

\begin{equation}
    Q^{new}(s,a) = (1-\alpha)Q(s,a) + \alpha (R(s,a) + \gamma Q(s',a^{*} )
    \label{eq:QL-main}
\end{equation} \\
where $\alpha$ and $\gamma$ are learning rate and discount factor respectively, $s'$ is the next state after taking action $a$ at state $s$ and $a^{*}$ is the action that results in the maximum Q-value of all state-action pairs on state $s'$:

\begin{equation}
    a^{*}=argmax_aQ(s',a)
    \label{eq:bestAction}
\end{equation}

In the above model, the agent (tUAV) needs to observe the rUAVs' geographical locations and their remaining battery levels. We assume our rUAVs remain in the same geo-cell in our considered area, therefore, only their battery status needs to be sent to the tUAV at each time step. Hence, our tUAV and rUAVs must have a light periodic signalling to exchange information.

Considering the discussed Q-Learning components, we follow Algorithm \ref{Q_Learning_Algorithm}, to obtain an optimal flying trajectory and recharging mechanism that maximizes the overall flying duration of all rUAVs. In this algorithm, the tUAV receives updated information of the rUAVs at each time step. The observed data including the tUAV current location indicate the current state. The agent makes a decision on movement either randomly or based on the Q-Learning policy. We employed $\epsilon$-greedy scheme for exploration in Algorithm \ref{Q_Learning_Algorithm} where the probability of selecting a random action decreases while the policy is being optimized through the iterations.  

\begin{figure*}
    \centering
    \includegraphics[width=0.75\textwidth]{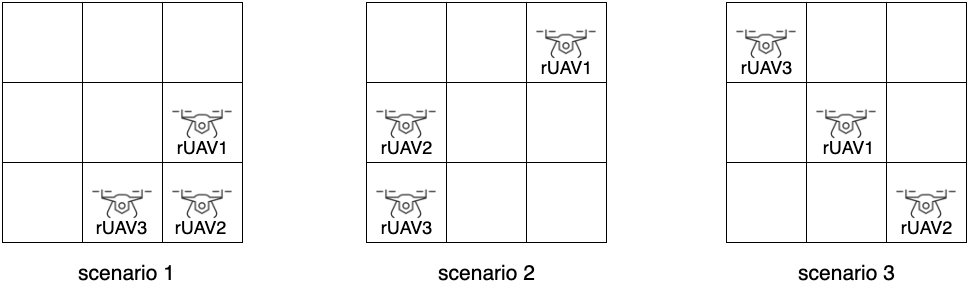}
    \caption{Considered simulation scenarios showing the positions of the rUAVs. The rUAVs are stationary in each scenario where the initial position of tUAV is randomized for each episode.}
    \label{fig:scenarios}
\end{figure*}

\section{Performance Evaluation}
\label{sec:perform}
\subsection{Simulation Setup}
\label{sec:simulation}

In our scenarios, we consider three rUAVs and one tUAV, located in an environment modeled as a 3 by 3 grid (Figure~\ref{fig:tUAVMove}, \ref{fig:scenarios}). To simplify our simulation design, we assume all tUAV and rUAVs can be located only at the center of these geo-cells. The tUAV sends the charging signal to all rUAVs at the same time and the closest rUAV receives the most amount of RF energy. Figure \ref{fig:tUAVMove} demonstrates possible flying directions (i.e., actions) of our tUAV when it's located in different geo-cells. Figure~\ref{fig:scenarios} illustrates three different scenarios that were considered in our simulations. For each scenario, we run our simulated Q-Learning model for 50000 episodes (i.e., iterations). In each episode, the tUAV starting cell is random. Also, the battery of each rUAV is initialized randomly between 60 and 100\,Watt-hour. The battery level is discretized to 5 levels as an observable parameter for Q-Learning. In the first episode, the exploration rate of Q-Learning is 1 and the trajectory of the tUAV is completely random. The tUAV location is updated at the start of each time step and the amount of consumed and received energy by each rUAV is also calculated at the end of each period. Consequently, the Q-Table is updated using equation~\eqref{eq:QL-main} when the new battery levels are observed. Since the harvested RF energy is less than rUAV energy consumption, rUAVs will be out of charge eventually. The episode continues until all rUAV batteries are discharged and they have to fly back to the terrestrial charging station. The exploration rate decreases linearly to zero after around 40000 episodes and then, the tUAV trajectory is fully based on Q-Table policy where at each observed state, the action with maximum Q-value is taken as the optimum decision using equation \eqref{eq:bestAction}. The simulation parameters are listed in Table~\ref{tab:sim_param}.

In order to evaluate our algorithm's performance, we consider the level of delivered wireless energy to the rUAVs and also simulate the following two baseline tUAV trajectory models as benchmark algorithms:

\begin{itemize}
    \item Random trajectory: all movements of the tUAV are random. Staying in the same geo-cell is also allowed in this model.
    \item Static hovering: The tUAV hovers in the central geo-cell for the entire episode.
\end{itemize}

\begin{algorithm}[t]
\caption{: Q-Learning}\label{Q_Learning_Algorithm}
\begin{algorithmic}[]
 \footnotesize
    \STATE  \ Initialize Q-Table to zero 
	\STATE  \ Initialize tUAV location
	\STATE  \ Set Exploration more than 1
    \STATE  \ Observe rUAVs locations
    \STATE  \ REPEAT 
    \STATE  \ \ \ \ Observe rUAVs Battery
    \STATE  \ \ \ \ Current state = (tUAV location,Observation)
    \STATE  \ \ \ \ Generate a random number $r$ $\in$[0,1]
    \STATE  \ \ \ \ IF r is less than Exploration
    \STATE  \ \ \ \ \ \ \ \  Update tUAV location by a random action
    \STATE  \ \ \ \ ELSE
    \STATE  \ \ \ \ \ \ \ \ Update tUAV location by the best action for current state
    \STATE  \ \ \ \ Calculate Revenue(consumed-received energy, out of charge rUAVs)
    \STATE  \ \ \ \ Update Q-Table by Revenue
    \STATE  \ \ \ \ Decrease Exploration 
    \STATE  \ CONTINUE
\end{algorithmic}
\end{algorithm}

\begin{table}[h]
\caption{Simulation Parameters}
    \centering
    \normalsize
    \begin{tabular}{|l|l|}
\hline
Q-Learning Component & Value \\ \hline
 Transmit power & 35 Watt \\ 
 Antenna gain & 25 dBi \\ 
 Cell side & 10 m \\ 
 Charging Wave Frequency & 25 GHz \\ 
 Learning rate & 0.4 \\ 
 Discount factor & 0.95 \\ 
 rUAV power consumption & 50 Watt \\ 
 rUAV battery capacity & 100 Watt-hour \\ 
 Time step & 20 Sec \\ 
 Revenue adjusting factor ($\mu$) & 100 \\ 
 Revenue adjusting factor ($\nu$) & -50 \\ \hline
    \end{tabular}
    \label{tab:sim_param}
\end{table}

\subsection{Results}
\label{sec:results}

\begin{figure}[t]
    \centering
    \includegraphics[width=0.49\textwidth]{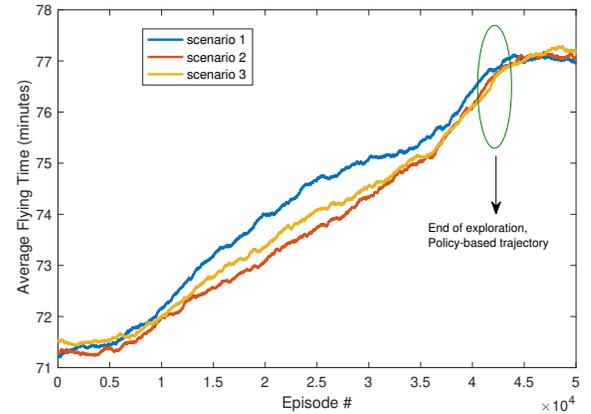}
    \caption{The average flying time of rUAV. The exploration rate is decreasing to the point that labeled on graphs.}
    \label{fig:3sResult}
\end{figure}

To demonstrate the exploration progress of Q-Learning Agent, the average flying time of rUAVs for three scenarios are shown in Figure~\ref{fig:3sResult}. This illustrates how the learning is progressing when at the start of the learning process, our Q-learning model takes totally random actions and gradually decreases this randomness by taking the Q-Table values into account. After around 40000 episodes, when the exploration rate drops to zero, the tUAV has a policy to find the best trajectory based on rUAV locations and their battery status. We find that flying time is increased from 71.5 to 77 minutes during the training session. 

\begin{figure}
    \centering
    \includegraphics[width=0.45\textwidth]{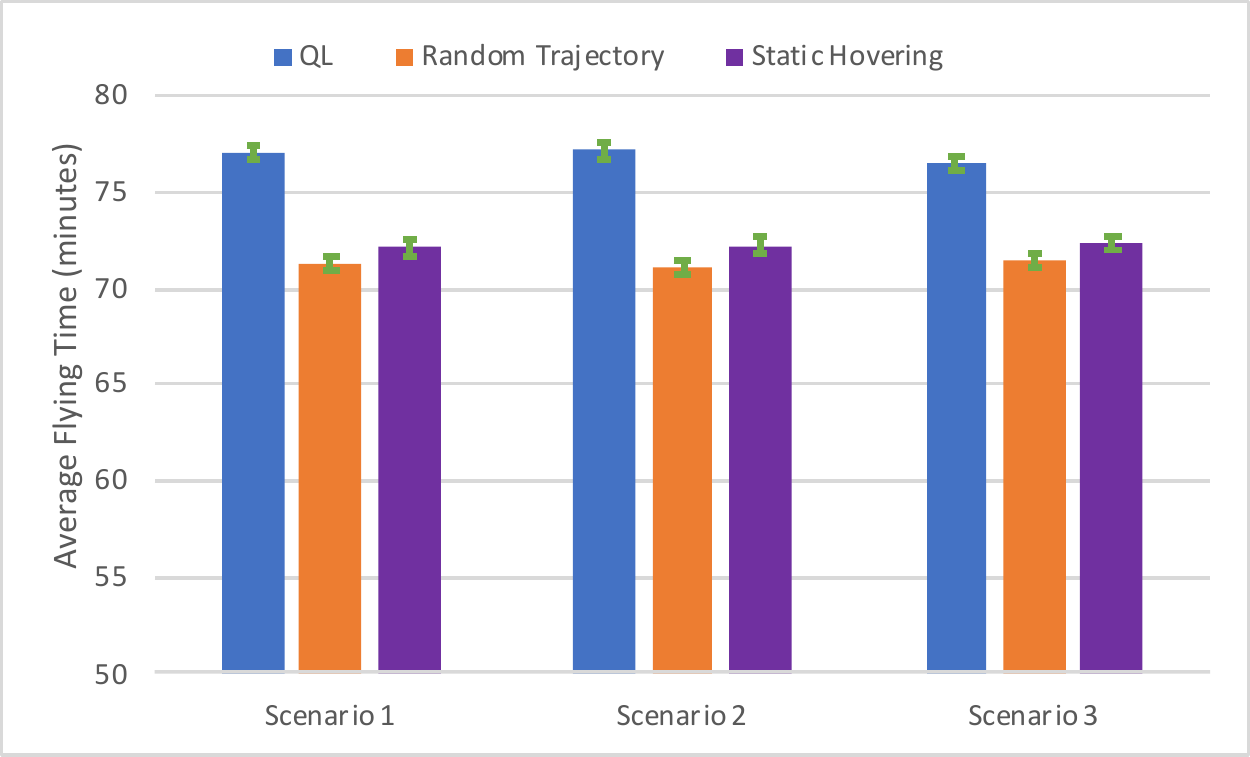}
    \caption{Comparison of flying time extension by positioning of the tUAV using Q-Learning and naive baselines.}
    \label{fig:barchart}
\end{figure}

In Figure~\ref{fig:barchart}, we compare how the Q-Learning based trajectory of the tUAV performs against that of two benchmark trajectories in terms of resulting average flying time of the rUAVs. As it is shown, the Q-Learning model significantly outperforms both methods since the tUAV makes decisions based on the knowledge obtained from previous iterations and aims to maximize the transferred energy and extend flying time of the rUAVs. The standard error bars also are shown on the chart which are mostly influenced by battery level randomness of each episode and are observed to be close for all schemes.

\begin{figure}
    \centering
    \includegraphics[width=0.45\textwidth]{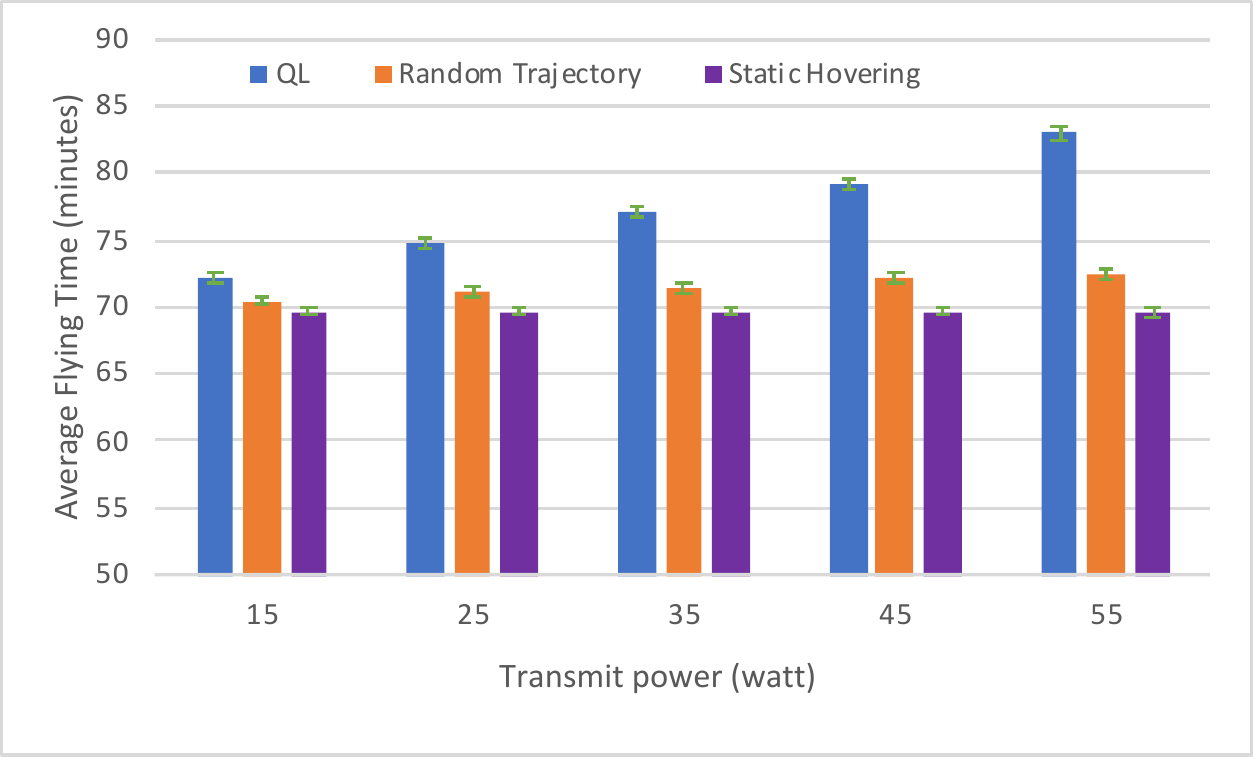}
    \caption{Comparison of flying time extension by positioning of the tUAV using Q-Learning and naive baselines for different transmit power. }
    \label{fig:power-bar}
\end{figure}

The energy transfer efficiency is limited mainly by distance and to increase the received energy, a more powerful electromagnetic wave can be emitted towards receivers. We repeat the Scenario-1's simulation for various transmission power. Figure~\ref{fig:power-bar} presents the average results. As it is illustrated, our Q-Learning based model is more effective than baseline positioning models when a higher amount of energy is emitted. For example, when the transmit power is 55~W, Q-Learning has an average flying time gain by 14\% and 19\% in comparison with the random trajectory and fixed hovering location of the tUAV at center of the map respectively. Currently, we have used a single tUAV in our solution, however, supporting multiple tUAVs would also enhance the level of received energy. This will require a multi-agent Q-Learning formulation. We will explore this multi-agent approach that supports multiple tUAVs in our future work.

\section{Discussion and Future Work}
\label{sec:conclusions}
We introduced the concept of using \textit{mobile, aerial} chargers for in-situ topping up of Hotspot UAV-BS battery using wireless power transfer. In order to enhance the level of received wireless power, we formulated the trajectory optimization problem of the aerial charger using MDP, and solved it using Q-Learning to maximize the flying time of energy thirsty UAV-BSs. Using simulation studies, we demonstrated that the Q-Learning based optimized trajectory of the aerial charger outperforms the benchmarking trajectories. Although our solution targeted Hotspot UAV-BSs that hover above fixed locations, it can be generalized for all applications where the power receiver UAVs (rUAVs) are hovering. Future work could consider other applications with mobile rUAVs. 

The Q-Learning is a dynamic programming method which can update the decision policy in an adaptive manner. In situations where the policy requires many experiences initially to achieve an acceptable performance, this training or exploration period can be estimated offline and loaded in the agent's software to avoid poor performance during the training phase. Thereafter, the agent which is the flying energy source (tUAV) in our studied problem, updates the policy using real environmental observations and experiences.

To avoid complexity, we chose a small scale scenario to demonstrate the usefulness of Q-Learning in the placement optimizations of the flying energy sources. However, a real scenario may be larger than what we have studied, where tens or hundreds of flying UAVs may \cite{agogino2012evolving} require in-situ charging in a large area and by multiple tUAVs. In such scenarios, basic Q-Learning may be too naive to meet the requirements and be practically implementable. For this purpose, more complex variations of Q-Learning need to be considered. In our future work, we will focus on methods such as Deep Reinforcement Learning \cite{lin1993reinforcement}, where the neural network is employed to support a Deep Q Network (DQN). This can support multi-agents and continuous values for observations and actions, while supporting a large scale environment.

\section*{Acknowledgement}
This work is supported by the Central Queensland University (CQU) Research Grant RSH5137.

\bibliographystyle{IEEEtran}
\bibliography{refs}
\end{document}